# Spline Collocation Method for Nonlinear Multi-Term Fractional Differential Equation

Dr. Vice-prof. *Choe Hui Chol*, Dr. Vice-prof. *Kang Yong Suk*

College of Mathematics, **Kim Il Sung** University

**Abstract** We study an approximation method to solve nonlinear multi-term fractional differential equations with initial conditions or boundary conditions. First, we transform the nonlinear multi-term fractional differential equations with initial conditions and boundary conditions to nonlinear fractional integral equations and consider the relations between them. We present a Spline Collocation Method and prove the existence, uniqueness and convergence of approximate solution as well as error estimation. The approximate solution of fractional differential equation is obtained by fractional integration of the approximate solution for fractional integral equation.

In the previous works[1-7], authors studied numerical methods of nonlinear one-term fractional differential equations with initial condition. Under some assumptions multi-term fractional differential equation was reduced to the system of one-term fractional differential equations. In the case of fractional differential equations with boundary conditions a finite difference method was studied under some assumptions for fractional order, and the existence of the positive solution was considered.[8] The existence of one-term fractional differential equations with integral boundary condition was studied.[11]

In this paper, we study nonlinear multi-term fractional differential equations with initial condition and boundary condition. We present a spline collocation method and prove existence, uniqueness of approximate solution and convergence with error estimation.

### 1. Preliminaries

**Definition 1** Let $\alpha > 0$.

$$I^\alpha y(t) = \frac{1}{\Gamma(\alpha)} \int_0^t (t-\tau)^{\alpha-1} y(\tau) d\tau$$ is called $\alpha$ order of fractional integral for $y(t)$.

This integral is defined for $y \in L(0, T)$.

**Definition 2** Let $\alpha > 0$, $n-1 < \alpha \leq n$. For $y \in C^n[0, T]$.

$$D^\alpha y(t) = \frac{d^n}{dt^n} \frac{1}{\Gamma(n-\alpha)} \int_0^t (t-\tau)^{n-\alpha-1} y(\tau) d\tau, \quad t \in [0, T]$$

defines Riemann-Liouville fractional derivative of order $\alpha$.

And $D_*^\alpha y(t) = \dfrac{1}{\Gamma(n-\alpha)} \int_0^t (t-\tau)^{n-\alpha-1} \dfrac{d^n}{d\tau^n} y(\tau) d\tau$ defines Caputo fractional derivative of order $\alpha$ for $y(t)$.

Many useful properties for the fractional derivative and fractional integral have been studied. We describe some of it for our purpose.

① $\alpha, \beta \geq 0$, $I^\alpha I^\beta = I^\beta I^\alpha$

② $\alpha, \beta \geq 0$, $I^\alpha I^\beta = I^{\alpha+\beta}$

③ $I^\alpha t^\gamma = \Gamma(\gamma+1)/\Gamma(\gamma+1+\alpha) \cdot t^{\gamma-\alpha}$, $\alpha > 0$, $\gamma > -1$, $t > 0$

④ $D^\alpha I^\alpha = I$, $\alpha \geq 0$

⑤ $D^\alpha t^\gamma = \Gamma(\gamma+1)/\Gamma(\gamma+1-\alpha) \cdot t^{\gamma-\alpha}$, $\alpha > 0$, $\gamma > -1$, $t > 0$

⑥ $I^\alpha D_*^\alpha f(x) = f(x) - \sum_{k=0}^{m-1} f^{(k)}(0) \dfrac{x^k}{k!}$, $m-1 < \alpha \leq m$, $m \in N$

⑦ $D_*^\alpha f(x) = D^\alpha \left( f(x) - \sum_{k=0}^{m-1} f^{(k)}(0) \dfrac{x^k}{k!} \right)$, $m-1 < \alpha \leq m$, $m \in N$

⑧ $f \in C[0, T]$, $I^\alpha f(0) = 0$, $\alpha > 0$

⑨ $f(x) = (x-a)^{\beta-1} \Rightarrow I_a^\alpha f(x) = \dfrac{1}{\Gamma(\alpha)} \int_a^x (x-t)^{\alpha-1} (t-a)^{\beta-1} dt = \dfrac{\Gamma(\beta)}{\Gamma(\alpha+\beta)} (x-a)^{\alpha+\beta-1}$

We consider the following equation.
$$D_*^\alpha x(t) = f(t, x(t), D_*^{\alpha_1} x(t), \cdots, D_*^{\alpha_n} x(t)), \ 0 < t \leq T$$

The Initial Value Problem of Nonlinear Multi-term fractional Differential Equation

Let $D_*^\alpha x(t) = f(t, x(t), D_*^{\alpha_1} x(t), \cdots, D_*^{\alpha_m} x(t))$, $0 < t \leq T$ \hfill (1)

$x^{(j)}(0) = x_j^0$, $j = 0, 1, \cdots, n-1$, $n-1 < \alpha \leq n$, $\alpha > \alpha_1 > \alpha_2 > \cdots > \alpha_n > 0$ \hfill (2)

where $x:[0, T] \to R$, $f:[0, T] \times R^{m+2} \to R$ are continuous.

When $x(t)$ satisfies (2.1), (2.2) and $D_*^\alpha x \in C[0, T]$, we say it a solution of (2.1), (2.2).

**Lemma 1** Let $x(t)$ be a solution of (1), (2) and $D_*^\alpha x(t) = y(t)$. Then $y(t)$ is a solution of the fractional integral equation

$$y(t) = f\left( t, \ I^\alpha y(t) + \sum_{k=0}^{n-1} x_k^0 \cdot \dfrac{t^k}{k!}, \ \left\{ I^{\alpha-\alpha_i} y(t) + D^{\alpha_i} \left( \sum_{k=n_i}^{n-1} x_k^0 \cdot \dfrac{t^k}{k!} \right) \right\}_{i=1}^m \right), \ n_i - 1 < \alpha_i \leq n_i \quad (3)$$

where $y \in C[0, T]$. And if $y \in C[0, T]$ is a solution of equation (2.3) and

$$x(t) := I^\alpha y(t) + \sum_{k=0}^{n-1} x_k^0 \cdot \dfrac{t^k}{k!}, \hfill (4)$$

then $x(t)$ is solution of a equation (1), (2).

**Proof** Let $x(t)$ be a solution of (1), (2). We set $y(t) := D_*^\alpha x(t)$. Applying $I^\alpha$ to two sides of equation and using property ⑥, we have the following expression.

$$I^\alpha y(t) = I^\alpha D_*^\alpha x(t) = x(t) - \sum_{k=0}^{n-1} x^{(k)}(0) \cdot \frac{t^k}{k!} = x(t) - \sum_{k=0}^{n-1} x_k^0 \cdot \frac{t^k}{k!}.$$

And by the property ⑦ $D_*^{\alpha_i} x(t) = D^{\alpha_i}\left(x(t) - \sum_{k=0}^{m_i-1} x^{(i)}(0) \frac{x^k}{k!}\right)$, $n_i - 1 < \alpha_i \leq n_i$, $i = \overline{1, m}$.

Hence,

$$D_*^{\alpha_i} x(t) = D^{\alpha_i}\left(I^\alpha y(t) + \sum_{k=0}^{n-1} x_k^0 \cdot \frac{t^k}{k!} - \sum_{k=0}^{n_i-1} x^{(k)}(0) \cdot \frac{t^k}{k!}\right) = I^{\alpha-\alpha_i} y(t) + D^{\alpha_i}\left(\sum_{k=n_i}^{n-1} x_k^0 \cdot \frac{t^k}{k!}\right). \quad (5)$$

Substituting (4), (5) to a expression (1), we have (3).

Next, let $y \in C[0, T]$ be a solution of (3). And we set

$$x(t) := I^\alpha y(t) + \sum_{k=0}^{n-1} x_k^0 \cdot \frac{t^k}{k!}, \quad n-1 < \alpha \leq n$$

Then

$$D_*^\alpha x(t) = y(t), \; 0 \leq n - \alpha < 1, \; y \in C[0, T]. \quad (6)$$

For $j = 0, 1, \cdots, n-1$,

$$D^j x(t)|_{t=0} = x^{(j)}(0) = D^j\left(I^\alpha y(t) + \sum_{k=0}^{n-1} x_k^0 \cdot \frac{t^k}{k!}\right)\bigg|_{t=0} = I^{\alpha-\alpha_i} y(t)|_{t=0} + x_j^0 = x_j^0 \quad (7)$$

$$D_*^{\alpha_i} x(t) = D^{\alpha_i}\left(x(t) - \sum_{k=0}^{n_i-1} x^{(k)}(0) \cdot \frac{t^k}{k!}\right) = I^{\alpha-\alpha_i} y(t) + D^{\alpha_i}\left(\sum_{k=n_i}^{n-1} x_k^0 \cdot \frac{t^k}{k!}\right). \quad (8)$$

Substituting (6)-(8) to an expression (3) we have (1). From (7) we have (2). □

From this lemma the problem that find a solution $x(t)$ of the equation (1)-(2) is reduced to the problem that find a solution $y(t)$ of the equation (3).

Next we consider the following initial value problem of the fractional differential equation.

$$D^\alpha x(t) = f(t, x(t), D^{\alpha_1} x(t), \cdots, D^{\alpha_m} x(t)), \quad (9)$$

$$x(0) = 0, \quad 1 > \alpha > \alpha_1 > \alpha_2 > \cdots > \alpha_m > 0 \quad (10)$$

We assume that $f$ is continuous on $[0, T] \times R^{m+1}$.

**Definition 3** If $x \in C(0, T) \cap L(0, T)$ satisfies (9)-(10) and $D^\alpha x \in C(0, T) \cap L(0, T)$, then we say that $x(t)$ is a solution of equation (9)-(10).

Let $C_r[0, T]$ is a space of functions $f \in C(0, T)$ with $t^r f \in C[0, T]$, where $0 \leq r < 1$. Norm on this space is defined as $\|f\|_r = \max_{t \in [0, T]} t^r |f(t)|$. Then $C_r[0, T]$ is Banach space. If $r = 0$ then $C_r[0, T] = C[0, T]$.

The following facts hold

Property-⑩

a) If $r < \alpha$, $f \in C_r[0, T]$ then $I^\alpha f \in C[0, T]$, $I^\alpha f(0) = 0$

b) If $f \in C[0, T] \cap L(0, T)$ then $D^\alpha I^\alpha f = f$

c) $C_r[0, T] \subset L(0, T)$

d) If $r < 1-\alpha$, $f \in C_r[0, T]$, $D^\alpha f \in C(0, T) \cap L(0, T)$ then $I^\alpha D^\alpha f = f$

**Lemma 2** Let $x(t)$ be a solution of equation (9)-(10) and $D^\alpha x \in C_r[0, T]$. Let $r < \alpha$, $0 \le r < 1$. If we set $D^\alpha x(t) = y(t)$, $y(t)$ is a solution of the fractional integral equation

$$y(t) = f(t, I^\alpha y(t), I^{\alpha-\alpha_1} y(t), \cdots, I^{\alpha-\alpha_m} y(t)). \tag{11}$$

And if $y \in C_r[0, T]$ is a solution of an equation (11) and $x(t) := I^\alpha y(t)$, then $x(t)$ is a solution of a equation (9)-(10).

The Boundary Value Problem of Nonlinear Multi-term fractional Differential Equation

We consider the following problem.

$$D^q_* x(t) = f(t, x(t), D^{q_1}_* x(t), \cdots, D^{q_n}_* x(t)), \ 0 < t < 1, \ 1 < q \le 2, \ 2 \ge q > q_1 > \cdots > q_n > 1 \tag{12}$$

$$A x(0) + B x'(0) = \eta_1, \ A x(1) + B x'(1) = \eta_2, \ A \ne 0 \tag{13}$$

**Lemma 3**[9] The general solution of the fractional differential equation $D^q_* x(t) = 0$ is $x(t) = c_0 + c_1 t + c_2 t^2 + \cdots + c_{n-1} t^{n-1}$, where $c_i \in R$, $i = 0, 1, \cdots, n-1$, $n-1 < q \le n$.

From lemma 3.1 there exists certain $c_i \in R$, $i = 0, 1, \cdots, n-1$ such that

$$I^q D^q_* x(t) = x(t) + c_0 + c_1 t + c_2 t^2 + \cdots + c_{n-1} t^{n-1}.$$

**Lemma 4** For $v \in C[0, 1]$ the unique solution of the boundary value problem $D^q_* x(t) = v(t)$, $0 < t < 1$, $1 < q \le 2$ and (13) is as follows.

$$x(t) = \int_0^1 G(t, s) v(s) ds + [(A(1-t) + B)\eta_1 + (-B + At)\eta_2]/A^2$$

$$G(t, s) = \begin{cases} \dfrac{A(t-s)^{q-1} + (B-At)(1-s)^{q-1}}{A\Gamma(q)} + \dfrac{B(B-At)(1-s)^{q-2}}{A^2 \Gamma(q-1)}, & s \le t \\ \dfrac{(B-At)(1-s)^{q-1}}{A\Gamma(q)} + \dfrac{B(B-At)(1-s)^{q-2}}{A^2 \Gamma(q-1)}, & t \le s \end{cases}$$

We consider the following equation instead of (12)-(13);

$$v(t) = f(t, I^q v(t) - b_1 - b_2 t, I^{q-q_1} v(t), \cdots, I^{q-q_m} v(t)) \tag{14}$$

**Lemma 5** For the solution of (14), $v \in C[0, 1]$

$$x(t) = I^q v(t) + \int_0^1 \left[ \dfrac{(B-At)(1-s)^{q-1}}{A\Gamma(q)} + \dfrac{B(B-At)(1-s)^{q-2}}{A^2 \Gamma(q-1)} \right] v(s) ds + \dfrac{1}{A^2}[(A(1-t) + B)\eta_1 + (-B + At)\eta_2]$$

is a solution of (12)-(13).

## 2. Spline Collocation Method for the Fractional Integral Equation

In the previous sections we transformed, fractional differential equations into fractional integral equations.

We consider equation (3). We can also consider similarly thing holds for (11), (14).

**Theorem 1** Let's suppose the following facts.

$f(t, z_0, z_1, \cdots, z_m)$ is continuous with respect to $t$.

$$|f(t, z_0, z_1, \cdots, z_m) - f(t, z_0'', z_1', \cdots, z_m')| \leq L\sum_{i=0}^{m}|z_i - z_i'|$$

$$\widetilde{L} := L \cdot \sum_{i=0}^{m} \frac{1}{\Gamma(\alpha - \alpha_i + 1)} < 1, \quad \alpha_0 = 0$$

Then there exists a unique solution of (3), $y \in C[0, T]$.

We consider spline collocation method for (3).

Let $\pi_N : 0 = t_0 < t_1 < \cdots < t_N = T$ be a partition of $[0, T]$. $t_i = ih, i = 0, \cdots, N$

By $S_1$ we denote the space of functions that are piecewise linear. We find $y_N \in S_1$ such that

$$y_N(t_i) = f\left(t_i, \ I^{\alpha}y_N(t_i) + \sum_{k=0}^{n-1} x_k^0 \cdot \frac{t_i^k}{k!}, \ \left\{I^{\alpha-\alpha_i}y_N(t_i) + D^{\alpha_i}\left(\sum_{k=n_i}^{n-1} x_k^0 \cdot \frac{t_i^k}{k!}\right)\right\}_{i=1}^{m}\right), \quad i = 0, \cdots, N.$$

(15)

Let $P_N$ be the projection in $C[0, 1]$ taking an arbitrary function $f \in C[0, 1]$ to its spline interpolation on the grid $\pi_N$, $f_N \in S_1$. We denote the right side of the equation (3) by $\varphi(y(t))$.

The problem that consider the uniqueness of solution of (15) is equivalent to the one of the equation $P_N y_N = P_N \varphi(y_N)$. Considering $P_N y_N = y_N$, $P_N y_N = P_N \varphi(y_N)$ can be written as $y_N = P_N \varphi(y_N)$. $\|P_N\| = 1$, so it follows that $y_N = P_N \varphi(y_N)$ has the unique solution under the assumption of theorem 1.

**Theorem 2** Let $y \in C[0, T]$ be a solution of (3) and $y_N \in S_1$ be an approximate solution. Then we have the estimation

$$\|y - y_N\| \leq c_k \|y^{(k)}\| \cdot h^k, \quad y \in C^k[0, T], \quad 1 \leq k \leq 2,$$
$$\|y - y_N\| \leq c_0 \omega(y, h), \quad y \in C[0, T].$$

Where $c_k, k = 0, 1, 2$ are constants independent to $y$, $h$ and

$$\omega(y, h) = \sup\{|y(t+\widetilde{h}) - y(t)| : t, \ t+\widetilde{h} \in [0, T], \ |\widetilde{h}| \leq h\}.$$